# UNDERSTANDING THE CHARACTERISTICS, BENEFITS AND CHALLENGES OF AGILE IT PROJECT MANAGEMENT: A LITERATURE BASED PERSPECTIVE

Godfred Yaw Koi-Akrofi[1], Joyce Koi-Akrofi[2] and Henry Akwetey Matey[3]

[1,3]Department of IT Studies, University of Professional Studies, Accra

[2]PMO Department, Vodafone Ghana

*ABSTRACT*

*The objectives of this study was to bring out the understanding of the concept of agile IT project management; what it is and what it is not. It was also aimed at comparing the pros and cons of both agile and traditional methods of IT project management in a typical industry setting; the challenges of going purely agile, and so on. It is purely a review of literature of peer reviewed papers sourced mainly from Google Scholar. It was revealed that agile outweigh the traditional methods in terms of benefits, but its implementation poses a lot of challenges due to a number of issues, paramount among them being organizational culture and empowerment of the project team. This has resulted in a number of industries sticking to the traditional methods despite the overwhelming benefits of agile. In another school of thought, the combination of the two paradigms is the way forward.*

## 1. INTRODUCTION

Seventeen individuals gathered at The Lodge at Snowbird Ski Resort in Wasatch Mountains in Utah on February 11-13, 2001 to speak, ski, relax and attempt to discover common ground — and, of course, eat.. What emerged was the Manifesto of Agile Software Development (Retrieved on 16/06/2019 at https://agilemanifesto.org/history.html).

Since the Agile Manifesto was released in 2001, agile approaches to software development projects have evolved significantly. According to many, particularly the writers of the manifesto, agility will become even more essential owing to the recognized reaction to modifications, while Bennekum and Van Hunt (in [1]) even argue that agile thinking is essential to 21st century achievement. Agile is a methodology in software application creation that anticipates the need for flexibility and applies a level of pragmatism to completed item delivery. Agile needs a cultural change in many businesses because it focuses on the smooth delivery of individual pieces or components of the software and not the entire implementation.

Agile project management (APM) is the outcome of the agile crusade for software development. APM is based on a 1986 document published for the Harvard Business Review by Hirotaka Takeuchi and Ikujiro Nonaka entitled "The New Product Development Game." In this article, the writers used rugby sport as a metaphor to describe the advantages of self-organizing teams in innovative product development and distribution [2]. Agile did not gain so much momentum until Jeff Sutherland and Ken Schwaber mentioned the first agile software development technique at the 1995 OOPSLA convention. After the rugby word that defines how teams form a circle and go to the ball to get it back into play, they called their fresh technique Scrum. At Easel Corporation in 1993, they first implemented this technique. In their book Agile Software Development with Scrum in 2002, Schwaber and Beedle wrote about their experiences, followed in 2004 by Schwaber's book Agile Project Management with Scrum, which included the job Schwaber had done with Primavera [2]. They discovered that traditional design methods were not appropriate for empirical, unpredictable and non-repeatable procedures while evaluating popular software development procedures [3].

Shane Hastie [4] describes how Agile differs from traditional techniques by focusing much more on team work, collaboration, and self-organization. One of the key to Agile's achievement is confidence, which must be present between the leader and the team as well as between the team members themselves. Some of the writers of the manifesto (Andrew Hunt, Arie Van Bennekum, and Jim Highsmith) also indicated in an interview with Bowles Jackson that the agile strategy is helpful for all kinds of projects and beyond. Asked directly if agility can be implemented outside IT projects, Hunt further claims that the agile strategy has little to do with software; instead, it is all about acknowledging and implementing feedback.

Van Bennekum adds that agile is holistic and applicable in company and life everywhere – he utilizes it as a notion wherever he is and for whatever he does. In every project facing uncertainty, the third co-author, Highsmith, thinks it is essential to use agile practices and principles. However, their foresight and efforts have not been confirmed by the exercise. Berger and Beynon-Davies [5]reviewed studies that examined the agile strategy in the last century and discovered that by 2009, almost all of them concentrated exclusively on IT projects; that year, one paper studied the agile strategy in product development projects.A year on, Doherty[6] studied agility initiatives in the field of e-learning, while Gonzalez [7] looked at the agile approach to intellectual property setting-up and management. In summary, only three out of thirty-three papers did not talk about IT projects [8, 9]

Despite the popularity of and much talked about agile IT project management, the traditional methods of IT project management are still being used by many companies. The question is then asked "Is agile IT project management implementation in firms just a mere talk or there are real challenges or problems to its full implementation?" This work in tends to deliberate on some of these issues to know the way forward from a literature based perspective.

The objectives of this study are therefore:

1. To help readers understand the concept of agile IT project management

2. To help IT project management practitioners understand the benefits of agile IT project management over other methods from a literature point of view

3. To help IT project management practitioners understand some of the challenges and problems associated with the full implementation of agile IT project management in firms from a literature point of view

4. To offer practical suggestions on how to maintain the momentum of agile IT project management full implementation in firms to make it a success after its introduction since 2001.

In the end, the study seeks to contribute to the general body of Knowledge in IT or IS project management, with agility in perspective.

## 2. LITERATURE REVIEW

### 2.1 What is Agile project management?

Agility, from the English Oxford living dictionaries, simply means "moving quickly and easily". Applying this same definition to project management, depicts or reveals the sharp move away from the traditional project management methodology like the waterfall, where we have a number of steps or processes to go through, one after the other till the last process is finished or executed before the project ends. These processes are such that one has to be done or finished before the next process can begin, which in most cases delay projects and also impedes the easy flow of the project. This write up focusses on software engineering or software development as an IT project, which is a deviation from all other project management fields like construction, and so on. In Agile project management, emphasis is placed on the earliest possible time to build the software and how easily the software is built for the customer as opposed to strict sticking to processes and complexity of designs and implementation procedures which in most cases result in delays.

According to Vikash Lalsing, Somveer Kishnah and Sameerchand Pudaruth [10], agile project management is a conceptual software engineering framework in which software is constructed within a comparatively brief span of time and has several iterations that result in stable software release. Agile is based on a set of values, according to the Agile Manifesto [11], focusing on client value, iterative and incremental implementation, intense cooperation, tiny integrated teams, self-organization, and tiny and constant improvements. Agile management is often said to work best with tiny teams. The perfect agile project team is tiny, collocated, communicate face-to-face on a daily basis and has an optimal team size that does not exceed nine individuals, according to Bustamante and Sawhney [12]. Agile management is often said to work best with tiny teams..

There are many distinct approaches to applying agile methods today, but underlying all the distinct agile movements are some fundamental ideas that turn traditional methodologies on their heads.The Agile Manifesto is a declaration articulating four (4) main values and 12 principles that software developers should use to guide their job according to the authors. The four key principles indicated in the "Manifesto for Agile Software[3] are:

(1) **Individuals and interactions over processes and tools:** With agile, the focus is on harnessing the skills of individuals into a formidable team resulting in constructive interactions for quick and easy delivery of projects. The focus shifts from the strict adherence to processes and tools as is the case for traditional methodologies.

(2) **Working software over comprehensive documentation:** With agile, the focus is on a software that works rather than focusing on elaborate documentation.

(3) **Customer collaboration over contract negotiation:** With agile, partnering with the customer is key to delivering quality software to the satisfaction of the customer than just insisting on contract terms, where deliverables are limited strictly to contract terms, and there is no room for flexibility.

(4) **Responding to change over following a plan:** With agile, there is a lot of flexibility, where a plan is not sacrosanct, but can be altered anytime there is the need for a change.

Again, the 12 principles articulated in the Agile Manifesto are:

1. ***Satisfying customers through early and continuous delivery of valuable work:*** Customers are satisfied with quality and early delivery of deliverables. Early delivery of deliverables does not mean quality is compromised.

2. ***Breaking big work down into smaller tasks that can be completed quickly:*** Tasks come together to form a project, and in this particular case, tasks are mostly deliverables that represent portions of the main project that can be deployed for use independent of the other tasks in executing process or yet to be executed.

3. ***Recognizing that the best work emerges from self-organized teams:*** Teams are small in size, and each member has a responsibility to ensure the success of the team. A team knows that the failure of one means the failure of the team. Jobs do not overlap in a team, and so failure to do one's job means a break in the link, and for that matter a looming failure awaiting the team. A team is not a group where one or two people will be doing the job on behalf of all the others, and all of them sharing in the success in the end, but pushing blame on the doers when there is failure. Agility places more emphasis on team work for excellent delivery where individual skills are harnessed.

4. ***Providing motivated individuals with the environment and support they need and trusting them to get the job done:*** The team members in an agile environment must be well motivated individuals who receive all the support they need to deliver quickly and effectively. Trust is key, and so the team members must trust themselves to deliver together in the first place, and the leader must also trust the team to deliver. Agility will not thrive in an environment where there is no trust.

5. ***Creating processes that promote sustainable efforts:*** Sustained efforts are needed to carry on to the end, and so the processes that are put in place must be agreed by all. Ground rules must be set by the team and must be agreed by all to ensure ownership of processes.

6. ***Maintaining a constant pace for completed work:*** Agility requires constant pace so that projects can be delivered within the shortest possible time. This is key, and without it, there is nothing like agility.

7. ***Welcoming changing requirements, even late in a project:*** Customer sophistication in terms of requirement changes is one of the main reasons why agility was given birth to. To deny customers of requirement changes in the course of the project delivery is like going back to the traditional methods of project management, and not agile project management employed.

8. ***Assembling the project team and business owners on a daily basis throughout the project:*** Constant monitoring on progress is essential to finishing the project quickly. This is because, with the daily constant meetings, all the challenges and problems can be ironed out quickly to allow for progress.

9. ***Having the team reflect at regular intervals on how to become more effective, then tuning and adjusting behavior accordingly:*** Regular team reflections will ensure that inefficiencies are weeded out as quickly as possible to ensure success at the end of the day.

10. ***Measuring progress by the amount of completed work:*** Progress is best measured by the amount of completed work so that strategies can be put in place to ensure the amount of work uncompleted are completed on time. Progress should not just be measured in the issues cleared, the number of change requests effected, and so on.

11. ***Continually seeking excellence:*** This is the way to go for agile IT project management; not being satisfied with excellence. Striving to do it better every opportune time is key.

12. ***Harnessing change for a competitive advantage:*** Agility should bring about change in the firm that goes a long way to give the firm an edge over its competitors in the industry.

Just as there are many kinds of projects, there are several distinct approaches to how agile methods can best be applied. Scrum, extreme project management, adaptive project management, and dynamic method of project management are some of the most significant ones. Of these, Scrum's overall model is most frequently used [3]. We take a close look at two of the techniques: SCRUM and Dynamic Systems Development Methodology (DSDM).

## 2.2 SCRUM as a technique of Agile Method

A Scrum is merely an agile, lightweight method for managing and controlling software and product development in fast-changing settings in terms of agile project leadership [3]. We should be clear about what Scrum isn't. Agile is Scrum is a prevalent misconception, and while Scrum is indeed agile, it is not the only technique employed for implementing agile principles. Scrum is just one of many agile product development methods. Other techniques include XP, Crystal, Driven Development feature, DSDM Atern, and so on [2]. All these techniques are in accordance with the Agile Manifesto and its related principles. A useful metaphor would be to think of Agile as an ice cream, while Scrum, XP, Crystal, etc. are all just distinct flavors, such as chocolate, strawberry, vanilla. All of them are agile; they are good, and many can be combined [2]. Scrum therefore, is simply an agile technique of delivering iterative and incremental products using frequent feedback and collaborative decision-making.

Agile project management Scrums are deliberately iterative, incremental procedures based on a team-based strategy. Given that systems are generally developing today in fluid settings and quickly changing environments, one of the main reasons for using an iterative method is to assist manage the chaos that may result from conflicting interests and needs within the project team. Iterative procedures are also used to assist improve communication, maximize collaboration, and safeguard the team from disruptions and impediments. Overall, the objective is to produce a product faster than traditional techniques.

### 2.2.1 Overview of SCRUM

The Scrum model consists of three main parts: roles, processes and artefacts [3].

   a. **ROLES**

   ❖ **The product owner**

According to Jeff Sutherland [13], the product owner:

> ➢ Defines the product characteristics or features, and also determines the release date and content
> ➢ Is responsible for product profitability (ROI)

- Prioritizes market value-based characteristics
- Can change or alter the features and priorities every 30 days
- Accepts or rejects outcomes or results of job

Basically, the product owner is a functional unit manager who understands what needs to be done to allow the project and how the construction sequence should advance [3].

### ❖ The SCRUM Master

The Scrum Master is the role that a team leader project manager has traditionally assumed. This individual is accountable for several factors, the most significant of which is perhaps the enactment of Scrum values and procedures and the removal of obstacles [3]. According to Jeff Sutherland (2005), the Scrum Master:

- Ensures full functionality and productivity of the team
- Enables close collaboration across all tasks and roles and removes obstacles
- Shields the team from external intrusions
- Ensures follow-up of the method or process. Calls for daily scrum, iteration review and meetings planning

### ❖ The SCRUM team

Typically, the Scrum team is a cross-functional team consisting of five to ten individuals working full-time on the project. The team is self-organizing, which has been interpreted in different ways, but most often implies that the role of leadership or management within the team is not fixed or permanent and changes or varies in process at the moment based on the requirements of the specific iteration (known as a sprint). It is essential to note that team affiliation or membership changes only between sprints. According to Jeff Sutherland [13], the Scrum team:

- Is Cross-functional, seven plus/minus two members
- Selects the objective of iteration and specify the outcomes of the job
- Has the right to do all within the limits of the project rules to achieve the objective of iteration
- Organizes itself and its work
- Demos work results to the Product Owner

### b. PROCESSES

The Scrum process has five major activities: the kickoff, the sprint planning meeting, the sprint, the daily Scrum, and the sprint review meeting.

**The Kickoff**

Similar to the sprint scheduling conference or planning meeting, the kickoff meeting is organized with the main distinction being that the group defines the high-level backlog for the project and the main project objectives or goals.

**The sprint planning meeting**

The sprint scheduling conference or planning meeting is a meeting at the start of each sprint between the Scrum team, the Scrum master, and the product owner. These meetings are made up of two components, which can take up to one day.

- ❖ PART 1: 2 Major activities:

    - ✓ 1st activity: The product backlog is defined by the group, which is basically a list of project demands or requirements.
    - ✓ $2^{nd}$ activity: The group determines the sprint objective or goal, which is the sprint's official outcome(s).

- ❖ PART 2

    The focus of the job is on establishing the sprint backlog in the second portion of the session or meeting.

**The Sprint**

The sprint can start once the sprint scheduling or planning meeting has taken place. In a traditional project, sprints vary from stages or phases in that sprints are restricted to a month-long iteration process in which the product's functionality is further developed.

Another differentiating factor is that no external influence should be permitted to interfere with the Scrum team's job during a sprint. This has several potential consequences with the most significant being that during a sprint it is not possible to change project demands.

Every sprint starts with a regular Scrum session or meeting in many projects, but not all. This meeting, typically no longer than 15 minutes in duration, is conducted daily between the Scrum master (who chairs the conference) and the Scrum team. Each member of the squad responds briefly to three issues in this session:

(1) What have you done since the last Scrum?

(2) Until the next Scrum, what are you doing?

(3) What's preventing you from working?

While it may not be obvious, the daily Scrum is not a problem-solving session and is not really intended to be a way to gather data about who is behind timetable or schedule (or what). Instead, the aim of the daily Scrum is both to monitor the team's progress and to allow team members to commit to each other and the Scrum master so that work can continue as expediently and unimpededly as possible. At the end of each sprint, the sprint evaluation meeting takes place. The functionality produced during the sprint is shown to the product owner during the conference. Perhaps the most distinctive characteristic of this conference from a traditional conference on project management is that it should be informal and not distracting team members [3].

    c. **THE ARTEFACTS**

This include the product backlog, the sprint backlog, and burn down charts.

- ➢ **Product backlog**

The Scrum Product Backlog is simply a list of all things to do within the project in the simplest definition. It replaces the artifacts of the traditional specification criteria. These items can be of a technical nature or in the form of user stories, for example, they can be user-centric. The Scrum

Product Backlog owner is the owner of the Scrum Product. The Scrum Master, the Scrum Team and other stakeholders are help contribute to have a comprehensive to - do list.
Working with a Scrum Product Backlog does not imply the creation and use of other artifacts by the Scrum Team is not permitted. A summary of the different user roles, workflow descriptions, user interface instructions, storyboards, or user interface prototypes could be examples for extra artifacts. These artifacts, however, do not substitute the Scrum Product Backlog, but complement its content and detail it.

During the Sprint Planning Meeting, the Scrum Product Owner utilizes the Scrum Product Backlog to define the team's top entries. Then the Scrum Team determines which objects during the coming sprint they can finish.

Each Scrum Product Backlog has certain properties that distinguish it from a simple to - do list:

- An entry in the Scrum Product Backlog always adds value to the customer
- Entries in the Scrum Product Backlog are prioritized and ordered accordingly
- The level of detail depends on the entry position in the Scrum Product Backlog.
- All entries are estimated
- Scrum Product Backlog is a living document
- The Scrum Product Backlog does not contain action items or low-level activities.

(Retrieved on 09/06/2019 at https://www.scrum-institute.org/The_Scrum_Product_Backlog.php)

> **Sprint backlog**

The Sprint Backlog is a collection of Product Backlog products chosen for the Sprint, plus a product Increment delivery schedule and a Sprint Goal achievement schedule. The Development Team's Sprint Backlog is a forecast of what functionality will be in the next Increment and the work required to deliver that functionality to a "Done" Increment. The Sprint Backlog shows all the job identified by the Development Team as needed to achieve the Sprint Goal. It involves at least one high priority process improvement recognized at the past Retrospective conference to guarantee continuous improvement. The Sprint Backlog is a plan with sufficient detail to allow the Daily Scrum to understand changes in progress. The Sprint Backlog is modified by the Development Team throughout the Sprint. During a Sprint, only the Development Team can alter their Sprint Backlog. The Sprint Backlog is a highly visible, real-time picture of the work planned by the Development Team during the Sprint, and it belongs to the Development Team alone. (Retrieved on 09/06/2019 at https://www.scrum.org/resources/what-is-a-sprint-backlog)

**Increment**

The Increment is the amount of all the Product Backlog items that were finished during a Sprint and the value of all prior Sprints increments. The new Increment must be "Done" at the end of a Sprint, which means it must be in useable condition and meet the definition of "Done" by the Scrum Team. An increment is a body of inspectable, done work that ends the Sprint with empiricism. The increment is a step toward a vision or goal. The increment must be in useable condition regardless of whether the Product Owner decides to release it (Retrieved on 09/06/2019 at https://www.scrum.org/resources/what-is-a-sprint-backlog)

> **Burn down charts**

In contrast to traditional project management, Scrum deliberately focuses on job performed by using burn-down charts. There are three kinds of burn-down charts frequently used:

- o **the sprint burn-downchart** documenting sprint progress,
- o **the release burn-down chart** documenting release progress, and
- o **the product burn-down chart** documenting general project progress.

A burn down chart's objective is to provide data in a way that is simple to understand. As such, each task is typically depicted by time (the display grid's x-axis) and length (y-axis). For example, the total backlog hours remaining in the sprint per day would be represented by a typical sprint burn down chart as an estimated amount of time left in the sprint [3].

Typically, the exceptional job is often on the vertical axis in a burn-down chart, with time along the horizontal. Predicting when all the work is done is useful.. The Development Team updates the Sprint Burn Down in the Daily Scrum and plots the rest of the day's job. For the following primary reasons, a burn down chart is almost a "must" instrument for a Scrum squad:

- ❖ monitoring the project scope creep
- ❖ Keeping the team running on schedule
- ❖ Comparing the planned work against the team progression

(Retrieved on 09/06/2019 at https://www.visual-paradigm.com/scrum/scrum-burndown-chart/)

## 2.3 Dynamic Systems Development Methodology (DSDM) as a technique of agile method

DSDM is an agile technique focusing on the entire lifecycle of the project. DSDM (officially known as the Dynamic System Development Method) was established in 1994 after RAD (Rapid Application Development) project executives sought more governance and discipline in this fresh iterative manner of working. The success of DSDM is due to the philosophy that "any project must be aligned with obviously specified strategic goals and concentrate on delivering actual company benefits early or quickly." They went out with eight principles to support this philosophy, enabling teams to keep focus and attain project goals.

The eight Principles of DSDM are:

- ✓ Focus on the business need
- ✓ Deliver on time
- ✓ Collaborate
- ✓ Never compromise quality
- ✓ Build incrementally from firm foundations
- ✓ Develop iteratively
- ✓ Communicate continuously and clearly
- ✓ Demonstrate control

DSDM is vendor-independent, includes a project's entire lifecycle and offers best practice advice for project delivery on time, in-budget delivery, with demonstrated scalability to tackle projects of all dimensions and for any business sector.

The DSDM promotes the use of a number of established standards or practices, including:

- Facilitated Workshops

- Modelling and Iterative Development

- **MoSCoW Prioritization:** MoSCoW is a technique for helping to comprehend priorities. The term *MoSCoW* itself is an acronym derived from the first letter of each of four prioritization categories (*Must have*, *Should have*, *Could have*, and *Won't have*). This is a method widely used in agile software development methods such as Scrum, Rapid Application Development (RAD) and DSDM, where a date is set to concentrate on the most significant demands. The MoSCoW method is a priority technique used in leadership, business analysis, project management and software development to achieve a common knowledge with stakeholders on the significance they attach to the fulfillment of each requirement ; it is also known as MoSCoW priority analysis or MoSCoW analysis. Although all demands are crucial, priority is given to early delivery of the biggest and most immediate company benefits. Initially, developers will attempt to implement all the *Must have, Should have and Could have* specifications or requirements, but *the Should and Could* specifications or requirements will be the first to be withdrawn if the shipping schedule or delivery timescale appears to be threatened. Compared to solutions such as High, Medium and Low, the simple English significance of priority classifications has significance in getting clients to better comprehend the effect of prioritizing.

    The categories are typically understood as:
    "Must have" Requirements marked as "must have" are critical to the present time box of delivery to make it a success. If even one requirement is not included, the delivery of the project should be regarded a failure (note: requirements can be downgraded from Must have, by contract with all appropriate stake holders; for instance, if fresh demands are considered more essential). *MUST* can also be considered an acronym for the Minimum Usable SubseT.

    *Should have* Requirements characterized as *Should have* are essential but not needed for delivery in the current delivery time box. While *Should have* requirements can be as essential as *Must have*, they often aren't as time-critical or there may be another way to meet the necessity so it can be held back to a future delivery time box.

    *Could have* Requirements characterized as *Could have* are desirable but not necessary, and may enhance user experience or customer satisfaction with low development costs. These will usually be included if time and resources permit.

    *Won't have* (this time) Requirements characterized as *Won't have* have been agreed by stakeholders as the least-critical, lowest-payback items, or not appropriate at that time. As a result, *Won't have* demands or requirements for the next delivery time box is not scheduled or programmed in the timetable or plan. *Won't have* criteria for integration in a subsequent time box are either dropped or reconsidered. (Note: the word *Would like to have* is sometimes used; however, the use is inaccurate, since this last priority obviously indicates that something is outside the scope of delivery).

- **Time boxing:** Time boxing allocates to each planned activity a fixed time period, called a time box. Many approaches to project management use time boxing. It is also used for individual use in a lower time frame to tackle private duties. A time box is a specified

International Journal of Software Engineering & Applications (IJSEA), Vol.10, No.5, September 2019period of time during which a job has to be achieved in agile software development. Time boxes are frequently used to handle the risk in developing software. The task of development teams is constantly to produce a releasable software enhancement or improvement, time boxed to a particular amount of weeks.

DSDM is designed to be easily tailored to traditional methods such as PRINCE2 ® or to complement other Agile approaches such as Scrum. DSDM is the backbone of Agile PM ® (Agile Project Management Accredited APMG) examination. This is the leading Agile Project management technique in the world and is increasing year after year in popularity.

**2.4 Main differences between Agile and Traditional Methods**

Agile is different from Traditional methods in many ways. The differences according to Hoda, Noble, and Marshall [14] are seen in the following areas: the development model, the focus of the project, the management of the project, Customer involvement, Developers, Technology, Product features, Testing, and Documentation. Table 1 below shows the comparative chart between Agile and Traditional methods:

**Table 1:** Comparative chart between Agile and Traditional methods

| CATEGORIES | TRADITIONAL | AGILE |
|---|---|---|
| Development Model | Traditional | Iterative |
| Focus | Process | People |
| Management | Controlling | Facilitating |
| Customer Involvement | Requirement gathering and delivery phases | On-site and constantly involved |
| Developers | Work individually within teams | Collaborative or in pairs |
| Technology | Any | Mostly Object Oriented |
| Product features | All included | Most important first |
| Testing | End of development cycle | Iterative and/or Drives code |
| Documentation | Thorough | Only when needed |

**Source:** Adapted from Hoda, Noble, and Marshall [14]

Realistically, in an IT project management environment, change requests as a result of frequent changes in customer requirements do occur, but this is not encouraged in traditional methods. In traditional methods, a lower phase or process group in the development life cycle has to be completed before moving on to a higher phase. This means that once requirements gathering and analysis is done, there is a very slim chance of changing requirements later in the development process, especially in the executing phase. This is resolved with agile methods. With agile methods, the focus is on the customer, and so requirements are welcomed and worked on periodically to ensure the satisfaction of the customer at the end of the day.

Aljaž Stare [9] also came out with four main differences between Agile methods and Traditional methods after an extensive research on agile methods. The differences are captured in the following headings:

- Requirements and specifications
- Project scheduling
- Team work, and
- The client's collaboration.

35

With Agility, the project team alone is not responsible for preparing requirements/specifications; it is the responsibility of both the client and the project team (www.agilemanifesto.org). Requirements are done in more detail at the start of the iterations, but are also considered and defined at the start of the whole project [15]. It includes an evaluation of functional significance [16], and revision or changes can be made upon the call of the clients or the project team members [17]. Moreover, in the project planning phase or later in the iteration planning phase, the least important features are eliminated [16].For traditional methods, almost the opposite of the above descriptions do happen; requirements are prepared by the project manager or team (but it must be emphasized that the project team gets most of the information from interactions with the client, and so it is not as if there are no collaborations between the team and the client. It is in the extent of collaboration and the preparation of the requirements document that the client is seen less), requirements are done at the very beginning of the entire project, requirements revision is not encouraged, even though can be done if it is of high priority necessity, but can only be effected with a change request, and lastly, all features are taken care of right from the beginning of the project to the end.

Traditional methods employ tools and techniques like the Gantt chart, Microsoft project, Critical path method, and so on to do project scheduling, where a project is normally broken down into several activities and arranged in order of importance depending on the dependencies, and executed in that order. In this case, overlapping is normally discouraged since an activity would have to be performed before another activity can be performed depending on the dependencies. With agility, the project is split into brief iterations which normally last no longer than eight weeks (Beck and Fowler in [18];[19]). The scheduling is actually done at the beginning of the project, while the scheduling of the iteration details is done by the team at the start of the iterations [20, 21]. Tactics for execution, assignments and performers are determined by the self-organized team [17, 20, 22], and before the alternatives or solutions are established, test processes are created [15].

With team dynamics, the traditional methods have the project manager, who is the CEO of the project delivery team, the subject matter experts which include the technical lead, and the other members like the change management specialist, public relations officer, and so on. The project and the whole team are managed by the project manager, and the project delivery or execution is done by the subject matter experts. On top of the project manager is the programmes manager or project owner, and on top of the programmes manager is the Steering committee or the Sponsor. These levels of authority also show escalation levels for issues resolution. With Agility, the team is accountable for the product's achievement on the market and not only for the project's efficient implementation, and constructive confrontation is a frequent way to find fresh thoughts and to solve problems [23]. Team members work in pairs [15, 19, 22), and they meet every day to discuss job outcomes, thoughts, issues and identify everyday duties [19]. Team members discuss and learn from their own errors on a regular basis, and following each iteration, the team will address the adequacy of the necessary assessments, methods and techniques, job errors and possible future improvements [20, 21].

Traditional methods do not have a representative of the client throughout the process groups; the client features most at the initial stages of the project, and is seen less after the initiation phase. With Agility, for further data, the client's representative is accessible 24/7 [17], one is a regular member of the team and participates actively every day [15]. The client representative Participates in test procedures implementation [16, 22], tests intermediate outcomes on a regular basis and reports insufficiencies and mistakes to the team [15], and proposes modifications (extra work and price, added value) and participates in their assessment.

Still on the differences between the traditional view and the agile perspective, Dybå and Dingsøyr [24] came out with a number of the quite different from the others discussed. Their work is summarized in Table 2 below:

**Table 2:** Differences between Traditional view and Agile perspective

| | TRADITIONAL VIEW | AGILE PERSPECTIVE |
|---|---|---|
| *Design process* | Deliberate and formal, linear sequence of steps, separate formulation and implementation, rule-driven | Emergent, iterative and exploratory, knowing and action inseparable, beyond formal rules |
| *Goal* | Optimization | Adaptation, flexibility, responsiveness |
| *Problem-solving process* | Selection of the best means to accomplish a given end through well-planned, formalized activities | Learning through experimentation and introspection, constantly reframing the problem and its solution |
| *View of the environment* | Stable, predictable | Turbulent, difficult to predict |
| *Type of learning* | Single-loop/adaptive | Double-loop/generative |
| *Key characteristics* | Control and direction. Avoids conflict. Formalizes innovation. Manager is controller. Design precedes implementation. | Collaboration and communication; integrates different worldviews Embraces conflict and dialectics. Encourages exploration and creativity; opportunistic. Manager is facilitator. Design and implementation are inseparable and evolve iteratively |
| *Rationality* | Technical/functional | Substantial |
| *Theoretical and/or philosophical roots* | Logical positivism, scientific method | Action learning, John Dewey's pragmatism, phenomenology |

**Source**: Adapted from Dybå and Dingsøyr [24]

The Traditional view adopts single-loop learning in the sense that the structure is fixed; there is no attempt by anyone to change the structure. The objective is to fix problems within the present organizational structure so that the system will function better. This makes the Traditional view rigid. The project manager has to work within prescribed structure or process; in other words, the project manager has to adapt to the existing structures and processes. On the other hand, the agile view is flexible, and employs double-loop learning. Double-loop learning is concerned with understanding the basis for the tasks being completed, rather than a more efficient process for completing them. It however, does not put aside entirely the existing structures and processes.
Again, Traditional view employs logical positivism, which argues that the only meaningful philosophical problems are those which can be solved by logical analysis. In other words, a scientific method or process must be in place to be used by project managers to undergo systems development or IT project management, rather than John Dewey's pragmatism which views or looks at knowledge as arising from an active adaptation of the human organism to its environment (The agile way).

Table 3 below shows the angle Sridhar Nerur, RadhaKanta Mahapatra, and George Mangalaraj [25] looked at the differences between Traditional and Agile Methods. The sub headings from which the differences were struck are good headings that can be used effectively to explain the advantages of both Traditional and Agile methods against each other, as well as the disadvantages against each other positing them in a particular context. These advantages and disadvantages are described in detail in the ensuing sections.

**Table 3:** Critical differences between Traditional and Agile Methods

| | Traditional | Agile |
|---|---|---|
| *Fundamental Assumptions* | Systems are fully specifiable, predictable, and can be built through meticulous and extensive planning. | High-quality, adaptive software can be developed by small teams using the principles of continuous design improvement and testing based on rapid feedback and change. |
| *Control* | Process centric | People centric |
| *Management Style* | Command-and-control | Leadership-and-collaboration |
| *Knowledge Management* | Explicit | Tacit |
| *Role Assignment* | Individual—favors specialization | Self-organizing teams—encourages role interchangeability |
| *Communication* | Formal | Informal |
| *Customer's Role* | Important | Critical |
| *Project Cycle* | Guided by tasks or activities | Guided by product features |
| *Development Model* | Life cycle model (Waterfall, Spiral, or some variation) | The evolutionary-delivery model |
| *Desired Organizational Form/Structure* | Mechanistic (bureaucratic with high formalization) | Organic (flexible and participative encouraging cooperative social action) |
| *Technology* | No restriction | Favors object-oriented technology |

**Source:** Sridhar Nerur, RadhaKanta Mahapatra, and George Mangalaraj [25]

## 2.5 Challenges and disadvantages associated with Traditional Methods and Benefits of Agile Methods

There are a number of challenges and disadvantages with traditional methods of systems development like the "waterfall" that make the benefits of agile development methods stand tall. Some of these disadvantages and challenges are outlined and discussed below:

i. Monolithic and slow [26].
ii. Process management is based on the volume of requirements: the higher the functionality of the requirements, the longer the delivery time of the project and the lower the flexibility and productivity of the process [26]. It is heavily dependent on initial requirements. However, if in any way these demands are defective, the project will be doomed.
iii. Management principles also contribute to multiple scheduling issues or problems, such as timely delivery, as it is difficult to predict the precise amount of job engaged at the planning phase of the project [26].

 iv. The enormous effort needed during a traditional project's scheduling stage is often so comprehensive that half (or more) of the project's funds are spent before any development job starts [3].
 v. Definitions of specifications or requirements are often so labor-intensive and prolonged that the project specifications have shifted or changed before implementation starts [3]
 vi. It is often rigid and resistant to change.
 vii. Only at the end is the entire item or product tested. If mistakes or errors are found late in the process, the remainder of the project may have been impacted by their presence.
 viii. The plan does not consider the changing or evolving requirements of a client throughout the entire project cycle.

In contrast to the above, Agile has the following advantages:

 i. It makes it possible to make modifications after the original scheduling phase. It follows the changes in customer demands or requirements.
 ii. Adding characteristics/features to maintain the product up-to-date with the industry's recent innovations is simpler.
 iii. Project priorities are assessed at the end of each sprint. This enables customers to add their feedback in order to finally get the product or item they want.
 iv. The end-of-sprint testing guarantees that the mistakes or errors in each cycle are captured.
 v. Agile practices give people some motivation to cooperate more than traditional techniques [27].

## 2.6 Challenges and disadvantages associated with Agile Methods

Despite the many advantages of Agile and the fact that the agile wave is still on, there are still some challenges implementing agile in organizations today. Most organizations cannot disregard the agile wave, but the implementation of agile methodologies is likely to present several difficulties for those steeped in traditional systems development. Some of the known disadvantages of agile development are as follows:

 i. This dynamic methodology is not suitable for processes that require a complex decision making of formal planning such as construction, manufacturing, military, health care system among others.
 ii. As the initial project does not have a definitive plan, the final product can be grossly different than what was initially intended.

There are three main reasons why Agile Methods fail in practice:

- *Insufficient experience with agile methods:* Agile looks simple, but requires a lot of experience in so many disciplines, especially social and managerial, to success in its usage. This is what many project management practitioners fail to understand. In terms of skills and experience of the team, Agile is rated high as compared to low experience of the traditional methods.
- *Little understanding of the wider organizational change needed:* Adopting Agile requires a complete shift from and sometimes the curtailment of the status quo, and this is understanding is not too clear to managers
- *Philosophy or culture of the company in contradiction with agile principles:* Any form of culture of an organization in contradiction to the Agile culture will not augur well for the thriving of Agile projects. In terms of company culture, Agile thrives better in a low management control environment as opposed to a highly management control environment for traditional methods for project management [28].

In other literature, from a practitioner point of view, there are 5 main disadvantages of Agile methodology:

- *Less foreseeability:* Developers are unable to quantify the full extent of required efforts for some software deliverables. This is principally true for larger products at the start of the development life cycle, and eventually leads to frustrations.
- *More Time and Dedication:* Testers, clients and developers must communicate with each other on an ongoing basis. This includes countless face-to-face discussions because they are the best way to communicate. Everyone engaged in the project must cooperate closely. Daily consumers need to be accessible on each stage for timely testing and sign-off so designers can mark it off as full before moving on to the next function. This may guarantee that the item meets the expectations of the customer, but it is burdensome and time consuming. This requires more time and energy from all involved.
- *Greater Developer and Customer Demands:* These principles involve close cooperation and comprehensive participation of users. Although it is an engaging and rewarding mechanism, it requires a great dedication to guarantee achievement throughout the project. Customers must be trained to assist in product development. Any absence of customer involvement will affect the quality and achievement of the software.
- *Lack of documentation required:* Because software requirements are explained just in time for development, there is less comprehensive documentation. This means they don't understand the information of certain characteristics or how they need to conduct when fresh employees join the team. This generates problems and misunderstandings.
- *Project falls off track readily:* To get started, this technique requires very little planning and assumes the requirements of the consumer are constantly evolving. You can see how this could restrict or slow the agile model with so little to go on. Then, if the feedback or communication from a consumer is not evident, a developer may concentrate on the incorrect development areas. It also has the capacity for creeping scope and becomes an ever-changing product (Accessed on 26[th] August, 2019 at https://www.inc.com/adam-fridman/the-massive-downside-of-agile-software-development.html).

Despite the popularity of the agile wave [11] the rate at which software projects are failing is still alarming. According to the Standish 2009 Chaos Report [29], the average success rate of all IT projects is 32 percent, 44 percent were challenged and the failure rate is 24 percent. The success rate of the project has declined from prior years, with only 32 percent being effective or successful, delivering them on time, on budget and with all the functionalities needed. 44% of the projects were questioned or challenged, meaning that either they were delivered late, over budget and/or with fewer originally planned functionalities. The remaining 24% of IT projects failed and had to be cancelled before completion and have never been used. In addition, according to Scott W. Ambler [30], achievement or success levels for agile projects are 60% effective, 28% challenged, and 12% failures. The success rate for Agile projects is 83 percent for small teams (less than eleven people), 70 percent for medium-sized teams (between eleven and twenty-five people), and 55 percent for large teams (more than twenty-five people). Based on these statistics, it is evident that agile team size directly affects the project's success rate.

Implementing agile technology comes with a lot of challenges for any organization. Employing the model of Sridhar Nerur, RadhaKanta Mahapatra, and George Mangalaraj [25], we discuss the challenges one after the other:

### i. Management and organizational challenges

This is categorized into the following sub headings: Organizational Culture, Management Style, Organizational Form, and Management of Software Development Knowledge, Reward Systems. If the management style is that of command-and-control as against leadership and collaboration, it will be very difficult for the organization to switch to agile development easily and quickly. The culture of an organization is the way of life of the organization (spoken and unspoken), and if the way of life of the organization is the mechanistic (bureaucratic with high formalization) as is the case for traditional system development, it will be very difficult for the it to adopt the organic (flexible and participative encouraging cooperative social action) in the case for agile systems development. Lack of management of software development knowledge can also pose a management or organizational challenge to the adoption of agile technology. This is because for the success with agile, one has to be abreast with even the management with the traditional methods, and the benefits with which agility comes to the table.

### ii. People

People challenges is categorized into the following: Effectively working in a team, high level of expertise, customer relationships — commitment, understanding or knowledge, closeness or proximity, trust, respect. When it comes to team dynamics with agile systems development, a number of questions do arise. Some of these questions are: what is the recommended size of the team to achieve high performance and excellent communication? What should be the level of expertise of the team members? How professional is the team when it comes to dealings with customers? As has already been established, research work by Scott W. Ambler [29] suggests that the smaller the team size, the higher the success rate of agile projects. Again, the findings stated in a study conducted by Ancona and Caldwell's [31] reveals that team size could have a direct effect on group efficiency and communication. Communication — either the absence of it, or miscommunication — is one of the root causes of project failure. Large teams are vehicles of bad communication inherently. This is based on the amount of geometrically, not linearly, growing channels of interaction [32]. There are three communication channels for a three-member team and ten for a five-member team, twice the amount of people. But a six-member team has as many as fifteen channels of communication. This is calculated by the communication channel complexity $(N \times (N - 1) / 2)$, where N is the number of persons in a team. The expertise of the agile team is crucial to the success of IT projects. Agile systems development depends so much on team cohesion, and so lack of a good team will spell doom for the projects undertaken by the team. A good team communicate well with each other and focus on goals and results. Everyone in the team contributes their fair share and they offer each other support. Team members are diverse and they have good leadership. They are organized and they have fun. Agile teams should have good attitudes towards customers because agile methods are customer centered as against traditional methods where customers only come in at the initial stages of the project management life cycle. The problem with very technical people is the difficulty in communicating effectively with customers. Communicating effectively and informally with customers to gather evolving requirements is a "must have" attribute if one wants to work in an agile team.

### iii. Process

With agile systems development, processes are no longer the strict adherence to stage by stage procedures and elaborate documentation and governance processes, but a flexible approach to project delivery. This makes agile simple and yet a discipline which requires knack. Lack of these expertise in itself is enough for agile to be rejected. Agile changes from process-centric to a feature-driven, people-centric approach. It is short, iterative, test-driven development that emphasizes adaptability, and managing large, scalable projects. Another issue is in the selection of an appropriate agile method.

### iv. Technology (Tools and Techniques)

This challenge is related to the suitability of current technology and instruments and fresh skill sets — refactoring, configuration management, and so on. It is very difficult, in fact almost impossible for an organization to change its technology, tools and methods to a new one when the current one they are using is working perfectly and achieving great results for them. This is why most organizations find it difficult to adopt agile after using the traditional methods for a long time and working perfectly for them. Coupled with the new skill sets that they have to acquire even makes them reluctant the more.

## CONCLUSION

Pedro Serrador and Jeffrey K. Pinto [33]evaluated the impact of agile use in organizations on two aspects of project success using a data sample of 1002 projects across various sectors and nations: effectiveness and general stakeholder satisfaction with organizational objectives. They also examined the moderating impacts of factors such as perceived quality of project vision / goals, complexity of the project, and experience of the project team. Their results indicate that both dimensions of project success are positively affected by Agile methods.

With all that have been discussed so far in this work relating to the benefits of both the traditional methods and agile methods and the their disadvantages, coupled with some empirical proofs, as is the case in the work of Pedro Serrador and Jeffrey K. Pinto [33], and industry information, it is clear that Agile methods are gaining and enjoying more popularity among industry players due to the overwhelming benefits over the traditional methods. Despite this, traditional methods continue to be the easy choice of many industries when it comes to IT projects and systems development because change is a difficult thing, and moreover a change that has to do with a total re-orientation of the status quo and the development of new skills all together. Some of these skills has to do with working effectively in small teams, being focused all the time, frequent reporting, presentation skills, customer involvement skills, understanding and ability to capture requirements quickly and also ability to modify requirements quickly, ability to take decisions quickly, ability to work without supervision, and so on. These re-orientation and skills gap normally do prevent industries to go all out for agile. Again, qualitative research with interviews with PMP and Prince2 and Agile communities revealed that for agility to be successful, two factors have to be addressed organizationally: organizational culture and the empowerment of the project team [34].

In another school of thought (views resulting from an interview with 5 selected seasoned or experienced IT programme managers), the way forward or the future is about harnessing and connecting the strengths of the two (Traditional and Agile).

As much as there is a need to churn out Products quickly, it is also very essential to deliver without chaos. Business Objectives for every Product require monitoring to ensure value delivery. Keeping focus does not mean use of strict Waterfall approach. A balance can be maintained by delivering Overall objects in sprints. The end goal can be delivered piece meal in a continuous cycle until total objects have been realized.

Furthermore, the question "why do we want to go agile?" is more important than just looking at the benefits of Agile, and wanting to implement it. Every organization should thoroughly consider why they need to be agile, how agile they really need to be, and what kind of agile projects they need to be? They can begin discussions and planning to decide the correct agile methodology after answering these questions [28].